\documentclass[
    ,final            
  ,numberedheadings 
  ]
  {aipproc}

\layoutstyle{6x9}

\usepackage{amsmath}
\usepackage{amsfonts}
\usepackage{amssymb}
\usepackage{latexsym}

 \let\b=\beta    
     \let\th=\theta   
\let\m=\mu    \let\n=\nu          \let\r=\rho
\let\s=\sigma

\newcommand{\be}{\begin{equation}}
\newcommand{\ee}{\end{equation}}
\newcommand{\bi}{\begin{itemize}}
\newcommand{\ei}{\end{itemize}}
\newcommand{\bea}{\begin{eqnarray}}
\newcommand{\eea}{\end{eqnarray}}

\newcommand{\EE}{\mathcal{E}}

\newcommand{\cO}{\mathcal{O}}
\newcommand{\cR}{\mathcal{R}}
\newcommand{\cS}{\mathcal{S}}

\newcommand{\half}{{\tfrac{1}{2}}}

\newcommand{\gb}{\bar{g}}

\newcommand{\p}{{\partial}}

\newcommand{\Db}{\bar{D}}

\newcommand{\Rb}{\bar{R}}

\begin{document}

\title{Four-derivative interactions in \\[1ex]
asymptotically safe gravity}

%

\author{Dario Benedetti}{
  address={Perimeter Institute for Theoretical Physics\\
31 Caroline St. N, N2L 2Y5, Waterloo ON, Canada\\
dbenedetti@perimeterinstitute.ca\\[1.5ex]}
}

\author{Pedro F. Machado}{
  address={Institute for Theoretical Physics and Spinoza Institute\\
Utrecht University, 3508 TD Utrecht, The Netherlands\\
P.Machado@phys.uu.nl \\[1.5ex]}
}

\author{Frank Saueressig}{
  address={Institute of Physics, University of Mainz\\
Staudingerweg 7, D-55099 Mainz, Germany \\
saueressig@thep.physik.uni-mainz.de \\[1.5ex]}
}

\begin{abstract}
We summarize recent progress in understanding the role of higher-derivative terms in the asymptotic safety scenario of gravity. Extending previous computations based on the functional renormalization group approach by including a Weyl-squared term in the ansatz for the effective action, further evidence for the existence of a non-Gaussian fixed point is found. 
The fixed point also persists upon including a minimally coupled free scalar field, providing an explicit example of perturbative counterterms being non-hazardous for asymptotic safety.
\end{abstract}

\maketitle


\section{Introduction}

Finding a consistent and predictive UV completion for gravity is one of the major challenges in theoretical high energy physics to date. As is well established, quantizing General Relativity perturbatively results in a non-renormalizable quantum field theory, necessitating the introduction of new counterterms at each order in the loop expansion. Including (non-supersymmetric) matter  does not improve this situation. For pure gravity, the renormalizability issue may be resolved by including the one-loop counterterms in the bare action and adding their quadratic part to the free propagator,
which leads to a  perturbatively renormalizable theory. This solution is also not satisfactory, however, as it is widely believed that the resulting theory  is  plagued by unitarity issues. Given such a situation, two possibilities open up naturally: the first is that gravity should be regarded as an effective field theory, which should be replaced by a more fundamental one with new degrees of freedom above a certain energy scale. The second is that the  ``non-renormalizability'' originates from the use of perturbation theory and may be overcome by a non-perturbative definition of quantum gravity.

The latter point of view naturally leads to Weinberg's asymptotic
safety conjecture \cite{Weinberg:1980gg} (see also the recent \cite{Weinberg:2009ca,Weinberg:2009bg} for a historical perspective and \cite{Niedermaier:2006wt,Percacci:2007sz,Reuter:2007rv} for reviews). This scenario is based on Wilson's modern viewpoint
on renormalization \cite{Wilson:1973jj} and envisages the existence of a non-Gaussian fixed point (NGFP) of the renormalization group (RG) flow with a finite number of ultraviolet-attractive (relevant) directions. For RG trajectories attracted to the NGFP in the UV (spanning the UV critical surface $\cS_{\rm UV}$), the fixed point ensures that the theory is free from uncontrollable UV-divergences, while the finite dimensionality of $\cS_{\rm UV}$ guarantees predictivity at all energy scales. This prescription generalizes the framework of perturbative renormalization, which is recovered in the case of the fixed point being Gaussian.

This scenario has been intensively investigated by means of the functional renormalization group equation (FRGE) for gravity \cite{Reuter:1996cp}, which has provided substantial direct evidence for such a NGFP \cite{Reuter:1996cp,Souma:1999at,Lauscher:2001ya,Reuter:2001ag,Litim:2003vp,Lauscher:2002mb,Codello:2008vh,Machado:2007ea}. Most of this evidence  arises from studying ``truncations'' of the exact flow equation, which entail projecting the RG flow of the full theory onto a finite dimensional subspace of the full theory space.
A profound test for the quality of these approximations consists in enlarging the truncation subspace and investigating the stability of the previous findings under this extension. Starting from the Einstein-Hilbert truncation, this program has been carried out by including terms with up to the eighth power of
the Ricci scalar \cite{Codello:2008vh,Machado:2007ea}, demonstrating a striking stability of the characteristic features of the NGFP.

The main caveat of these $f(R)$-type truncations, however, is that they omit the four-derivative propagator for the helicity two states, which could, in principle, have a major impact on the asymptotic safety scenario. An a priori argument highlighting the importance of these terms is that they feature as non-renormalizable counterterms in the perturbative quantization of General Relativity.\footnote{In the case of pure gravity, these  counterterms vanish on-shell, so that the first divergence comes with the Goroff-Sagnotti term at two loops. However, once a free scalar field is included, these counterterms are non-trivial also on-shell, signaling the perturbative non-renormalizability of the theory.} In \cite{us1,us2} we closed this gap by explicitly including such higher-derivative terms in the truncation subspace. Our results confirm the existence of the NGFP already observed within other truncations, while at the same time underlining the importance of the new terms for the gravitational RG flow, as the effects originating from including these higher-derivative terms are significantly larger than the ones observed when varying the gauge-fixing term or cutoff-scheme within an $f(R)$-type truncation. In the sequel, we review the setup and main results of our work, referring to \cite{us1,us2} for technical details.

\section{Basics: the functional renormalization group equation for gravity}

\begin{figure}
  \includegraphics[width=0.90\textwidth]{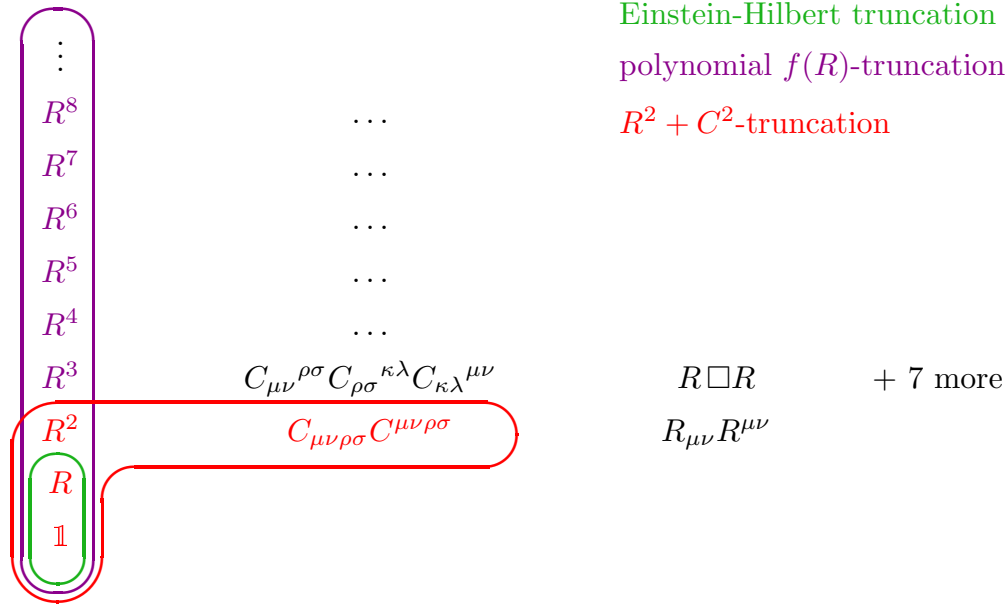}
  \caption{Illustration of the systematic exploration of the theory space of pure gravity employing truncations. The colored circles indicate the interaction monomials contained in the various truncation ans{\"a}tze for $\bar{\Gamma}_k[g]$. The $R^2 + C^2$ truncation thereby captures the non-perturbative RG flow 
of the two linear combinations \eqref{u2u3}, built from the three four-derivative couplings of higher-derivative gravity.
All truncations have confirmed the existence of a non-trivial UV fixed point of the gravitational RG flow.}
\label{Fig.1}
\end{figure}

A powerful tool for studying the non-perturbative properties of the gravitational RG flow is 
the FRGE for the effective average action $\Gamma_k$ (Wetterich equation) \cite{Wetterich:1992yh} adapted to gravity \cite{Reuter:1996cp},
\be\label{FRGE}
\p_t \Gamma_k[\Phi, \bar{\Phi}] = \half {\rm STr} \left[ \left( \frac{\delta^{2} \Gamma_k}{\delta \Phi^A \delta \Phi^B } + \cR_k \right)^{-1} \, \p_t \cR_k  \right]\, .
\ee
Here, $\Phi$ and $\bar{\Phi}$ respectively denote the physical fields and their background value,  $t = \log(k/k_0)$, and {\rm STr} is a generalized functional trace which includes a minus sign for ghosts and fermions, and a factor two for complex fields. Furthermore, $\cR_k$ is a matrix valued IR cutoff, which, at the level of the path-integral, provides a $k$-dependent mass-term for fluctuations with momenta $p^2<k^2$. Thanks to the factor $\p_t \cR_k$, the r.h.s.\ of the flow equation is finite and peaked at momenta $p^2 \approx k^2$, so that an additional UV-regulator becomes dispensable. The background covariance of $\Gamma_k[\Phi, \bar{\Phi}]$ is ensured via the background-field method, decomposing the physical metric $g_{\mu\nu}=\bar{g}_{\mu\nu} + h_{\mu\nu}$ into the background metric $\bar{g}_{\mu\nu}$ and arbitrary fluctuations $h_{\mu\nu}$ (and similarly for the other fields).

The FRGE realizes the idea of the Wilsonian renormalization group on the space of all action functionals, built from $\Phi$, compatible with the symmetries of the theory. 
It describes the dependence of $\Gamma_k[\Phi, \bar{\Phi}]$ on the coarse-graining (or renormalization group) scale $k$, essentially ``integrating'' or ``coarse graining'' 
over quantum fluctuations with momenta $k^2 \le p^2 \le \Lambda_{\rm UV}$. Its solutions interpolate between the ordinary effective action $\Gamma \equiv \Gamma_{k \rightarrow 0}$ and an initial action $\Gamma_\Lambda$ at the UV cutoff scale, which  in the limit $\Lambda_{UV} \rightarrow \infty$ essentially reduces to the bare action (see \cite{Manrique:2008zw} for more details).

This setup turns out to be ideal for investigating the asymptotic safety scenario,
providing a method for studying the gravitational RG flow non-perturbatively. 
Making an expansion
\be\label{expand}
\Gamma_k[\Phi, \bar{\Phi}] = \sum u_i(k) \, \cO_i[\Phi, \bar{\Phi}]\,,
\ee
where $\cO_i[\Phi, \bar{\Phi}]$ is a basis of all possible interaction monomials, eq.\ \eqref{FRGE} provides the $\beta$-functions for the (essential) dimensionful couplings $u_i(k)$. When analyzing the properties of the RG flow, however, it is more convenient to switch to the dimensionless coupling constants $g_i = k^{-d_i} u_i$, where $d_i$ is the mass-dimension of $u_i$. This results in autonomous $\beta$-functions $\p_t g_i = \beta_{g_i}(\{ g_j\})$. 
The central ingredient of the asymptotic safety program is a NGFP $\{\b_{g_i}(\{ g_j^*\})=0, g_i^* \not = 0\}$ of these 
$\b$-functions, which provides a well-defined
and predictive continuum limit.
 
Around any FP, the linearized RG flow, $\p_t g_i = {\bf B}_{ij} (g_j - g^*_j)$, is governed by the stability matrix 
${\bf B}_{ij} = \left. \p_j \beta_i \right|_{\{g_i^*\}}$.
Defining the stability coefficients $\th_i$ as minus the eigenvalues of $\bf B$, the relevant (irrelevant) directions are associated to the eigenvectors corresponding to stability coefficients with a positive (negative) real part. Renormalizable theories correspond to
RG trajectories which are attracted towards the fixed point in the UV. These trajectories define the UV-critical surface of the fixed point, whose dimension is given by the number of stability coefficients with ${\rm Re}\theta > 0$. Collecting further evidence for the existence of such a fixed point and determining the dimension of its UV-critical surface are two of the main tasks of the asymptotic safety program.  

The main shortcoming of the FRGE is that it cannot be solved exactly and,
in order to extract physics from it, one therefore has to resort to approximations.
One possibility is, of course, perturbation theory. In the one-loop approximation, where $\Gamma_k$ under the STr is replaced by the $k$-independent bare action, 
this leads upon integration to the usual non-renormalizable logarithmic divergences \cite{Codello:2008vh}. 
 
Going beyond perturbation theory, a standard approximation scheme is the truncation of the RG flow, where the flow of the full theory is projected onto a subspace spanned by a finite number of interaction monomials by restricting eq.\ \eqref{expand} to 
a finite subset of the $\cO_i[\Phi, \bar{\Phi}]$. Substituting this ansatz into the FRGE, this technique then allows one to extract the $\beta$-functions for the $g_i$ contained in the ansatz without having to rely on an expansion in a small parameter and while keeping contributions to all loop-orders, thus attaining non-perturbative results. In the next sections, we will summarize the central properties of the gravitational RG flow studied within a novel higher-derivative truncation employing this technique. 

%
\section{Setup: Gravity in the higher-derivative truncation}

When computing the $\beta$-functions within a truncation of \eqref{FRGE}, the main limitation arises from  our ability to evaluate the functional traces. A (still very general class of) truncations consists in making the ansatz
\be
\Gamma_k[\Phi, \bar{\Phi}] = \bar{\Gamma}_k[\Phi] + \widehat{\Gamma}_k[\Phi - \bar{\Phi}, \bar{\Phi}] + 
S^{\rm gf}[g, \gb] + S^{\rm gh}[g, \gb, {\rm ghosts}] + S^{\rm aux} ,
\ee
which restricts the ghost action $S^{\rm gh}[g, \gb, {\rm ghosts}]$ to the classical one. In this decomposition, $\bar{\Gamma}_k[\Phi]$ depends on the physical fields only, $S^{\rm gf}$  denotes the {\it classical} gauge-fixing term, and $S^{\rm aux}$ collects the contribution of auxiliary fields introduced to exponentiate the Jacobi-determinants arising from field-decompositions. The $\widehat{\Gamma}_k$ encodes the deviations from $\bar{\Gamma}_k[\Phi]$ for $\Phi \neq \bar{\Phi}$. By definition, it vanishes for $\Phi = \bar{\Phi}$, and captures, e.g., the quantum corrections to the gauge-fixing sector. ``Bi-metric'' truncations with non-trivial $\widehat{\Gamma}_k[\Phi - \bar{\Phi}, \bar{\Phi}]$ have recently been investigated in the conformally reduced setting in  \cite{Manrique:2009uh}, while a first step towards understanding the quantum corrections in the ghost sector has been undertaken in \cite{Eichhorn:2009ah}. Both works also confirmed the existence of the NGFP, lending additional support to the asymptotic safety scenario.

In the sequel, we shall follow most of the FRGE-literature and focus on truncations of the form
\be\label{EAA}
\bar{\Gamma}_k[\Phi] = \Gamma_k^{\rm grav}[g] + n_s \,  \Gamma^{\rm matter}[g, \phi] \, , \qquad
\hat{\Gamma}_k = 0 \, ,
\ee
where $n_s=0,1$ has been introduced to conveniently label the pure-gravity and matter-coupled cases.
Here, $\Gamma_k^{\rm grav}$ is the gravitational part of the effective average action, in which we include all interaction terms up to fourth order in the derivative expansion,
\be \label{ansatz}
\Gamma_k^{\rm grav}[g] = \int d^4x \sqrt{g} \left[ \frac{1}{16 \pi G_k} (-R + 2 \Lambda_k) - \frac{\omega_k}{3 \sigma_k} R^2 + \frac{1}{2 \sigma_k}  C^2   + \frac{\theta_k}{\s_k} E\right] \, ,
\ee
with $C^2 \equiv C_{\m\n\r\s} C^{\m\n\r\s}$ being the square of the Weyl tensor,
and $E = R_{\m\n\r\s} R^{\m\n\r\s} - 4 R_{\m\n} R^{\m\n} +  R^2$, the integrand of the Gauss-Bonnet topological invariant. The  $k$-independent action for the minimally coupled free scalar field is in turn given by
\be\label{matter}
\Gamma^{\rm matter}[g, \phi] = \half \, \int d^4x \sqrt{g} \,  \, g^{\mu\nu} \, \p_\mu \phi \, \p_\nu \phi\, .
\ee
Lastly, the gauge-fixing is implemented via 
\be\label{S:gf2}
S^{\rm gf} = \frac{1}{2} \int d^4 x \sqrt{\gb} \, F_\mu \, Y^{\mu\nu} F_\nu \, , \quad
F_\mu = \Db^\nu h_{\mu \nu} - \tfrac{1}{4} \Db_\mu h \, , \quad Y^{\mu \nu} =  \beta \, \gb^{\mu \nu} \, \Db^2 \, .
\ee
 where we take the limit $\beta \rightarrow \infty$ under the trace of the flow equation.

When computing the functional traces, the background field formalism ensures that we are free to choose a particular class of background metrics to simplify our computation, e.g., by making the resulting operator expressions amenable to current heat-kernel techniques. Dealing with truncations of the form \eqref{EAA} thereby requires 
a class of backgrounds which is generic enough to disentangle the coefficients multiplying $R^2$ and the tensorial terms, and, most importantly, simple enough to avoid the appearance of non-minimal 
higher-derivative differential operators. While the maximally symmetric backgrounds previously used are insufficient in the former respect, a generic compact Einstein background $\EE$ (which we take without Killing or conformal Killing vectors and without boundary for simplicity), satisfying
$\Rb_{\m\n}=\tfrac{\Rb}{4} \, \gb_{\m\n}$, 
is better suited to meet both criteria. In this case, all differential operators appearing in our traces can be organized in Lichnerowicz form, 
\be\label{def:LL}
\begin{split}
\Delta_{2L} \phi_{\mu\nu} & \equiv -D^2 \phi_{\mu\nu} - 2 R_{\mu\,\,\,\nu}^{\,\,\,\alpha\,\,\,\beta} \phi_{\alpha\beta}\,,\quad
\Delta_{1L} \phi_\mu  \equiv \left[ -D^2  - \tfrac{1}{4}  R \right] \,  \phi_\mu\,,\quad
\Delta_{0L} \phi  \equiv -D^2 \phi\, ,
\end{split}
\ee
and the trace evaluation may then be carried out utilizing the early time expansion of the heat kernel adapted to such operators, as detailed in Appendix~B of \cite{us2}.
These backgrounds allow us to distinguish two of the three higher-derivative couplings, and hence determine the non-perturbative $\beta$-functions of the linear combinations
\be \label{u2u3}
u_2 = - \frac{\omega_k}{3\sigma_k} + \frac{\theta_k}{6\s_k} 
\, , \qquad u_3 = \frac{1}{2\s_k} + \frac{\theta_k}{\s_k} \, .
\ee 
The generalization of the background metrics from maximally symmetric to Einstein provides the crucial ingredient for investigating the non-perturbative features of the gravitational RG flow including higher-derivative tensorial operators.

\section{Results: non-perturbative fixed point structure of the RG flow}

Before embarking on the analysis of the non-perturbative $\beta$-functions, we first verify that our flow equations correctly reproduce the known one loop $\beta$-functions for higher-derivative gravity 
obtained via dimensional regularization \cite{Fradkin:1981iu,Avramidi:1985ki}. For this purpose, we ``switch off'' the RG improvement of the FRGE, by neglecting the running of the coupling constants on its r.h.s.,\ and expand the resulting expressions to first order in $\sigma$. Following this route, we then arrive at the following, universal (i.e., $\cR_k$-independent) one-loop result:
\be\label{proj1l}
\beta_\sigma = - \tfrac{1}{(4\pi)^2} \tfrac{413}{90} \sigma^2 \, , \qquad 
\beta_\omega = \tfrac{1}{(4\pi)^2} \tfrac{317 - 1726 \omega - 600 \omega^2}{180} \sigma \, .  
\ee
Note that these are precisely the known one-loop $\beta$-functions of \cite{Fradkin:1981iu,Avramidi:1985ki} projected onto an Einstein background.
 
When analyzing the full, non-perturbative result, it is convenient to work with the dimensionless couplings
\be\label{dimless}
g_0 = \frac{\Lambda_k}{8 \pi G_k}k^{-4} \; , \quad g_1 = -\frac{1}{16 \pi G_k} k^{-2} \; , \quad g_2 = u_2 \, , \quad g_3 = u_3 \, ,
\ee
together with the dimensionless Newtons constant $g_k = G_k \, k^2$ and cosmological constant $\lambda_k = \Lambda_k \, k^{-2}$.
The corresponding $\beta$-functions in our matter-coupled four-derivative truncation \eqref{EAA} were presented in \cite{us2}.
Owed to their intricate structure, these $\beta$-functions can only be analyzed numerically. Furthermore, their explicit evaluation implies specifying the particular profile of the IR cutoff function.
In the following, all results are given for the optimized cutoff \cite{Litim:2001up}, 
whose scalar part takes the form $R_k(p^2) = (k^2 - p^2) \theta(k^2 - p^2)$.

The fixed point structure of these $\beta$-functions exhibits, first, two Gaussian fixed points
\be
g^* = 0  \, , \qquad \lambda^* = 0 \, , \qquad \sigma^* = 0 \, , \qquad 
\omega^*_{1,2} = - \tfrac{1}{120} \left(90 \pm \sqrt{15708 - 101 n_s} \right) \, ,
\ee
with stability properties given by the following eigensystem,
\be
\begin{array}{llll}
\theta_1 = 2 \, , \qquad & V_1 =  \{ 1,0,0,0 \}^{\rm T} \, , \qquad  &
\theta_2 = -2 \, , \qquad & V_2 =  \{ \frac{2+n_s}{16 \pi},1,0,0 \}^{\rm T} \, , \\
\theta_3 = 0 \, , \qquad & V_3 =  \{ 0,0,1,0 \}^{\rm T} \, , \qquad  &
\theta_4 = 0 \, , \qquad & V_4 = \{ 0,0,0,0 \}^{\rm T} \, .
\end{array}
\ee
 These GFPs correspond to the free theory. Their stability coefficients are thus given by the canonical mass dimension of the relative (dimensionful) couplings and the eigendirection associated with Newton's constant are UV repulsive, while those directions associated with the higher-derivative couplings are marginal.  The latter  eigendirections were found to be UV-attractive when going beyond the linear approximation, in accordance with the one-loop results \cite{Codello:2006in}.

Most importantly, both the pure gravity and the matter-coupled higher-derivative truncation also give rise to a NGFP with positive Newton's and cosmological constant. Its corresponding position and stability coefficients are
\be\label{FPpos}
\begin{array}{llllll}
n_s = 0: \; \; &  g_0^* = 0.00442 \, , \; \; & g_1^* = -0.0101 \, , \;\; & g_2^* = 0.008 \, , \; \; & g_3^* = -0.0050 \, , 
 \\
n_s = 1: \; \; &  g_0^* = 0.00438 \, , \; \; & g_1^* = -0.0087\, , \;\; & g_2^* = 0.010 \, , \; \; & g_3^* = -0.0043 \, , 
\end{array}
\ee
\be\label{FPstab}
\begin{array}{lllll}
n_s = 0: \;\; & \theta_0 = 2.51 \, , \;\; & \theta_1 = 1.69 \, , \;\; & \theta_2 = 8.40 \, , \;\; & \theta_3 = -2.11 \, , \\
n_s = 1: \;\; & \theta_0 = 2.67 \, , \;\; & \theta_1 = 1.39 \, , \;\; & \theta_2 = 7.86 \, , \;\; & \theta_3 = -1.50 \, .  
\end{array}
\ee
A first thing to notice is that the location of the fixed point in the $\{g_0,g_1\}$ plane is very close to that of the corresponding NGFP found within previous truncations. A second very important observation is that the inclusion of the scalar field does not change the qualitative picture, only minimally affecting the numerical values.

One salient difference with the Einstein-Hilbert case is the fact that all stability coefficients are now real. This is in agreement with the one-loop results of \cite{Codello:2006in}, but it is surprising that, unlike in the Einstein-Hilbert case, the transition from the one-loop to the non-perturbative treatment does not give rise to complex eigenvalues. We can trace this difference to the contribution of the $C^2$ terms coming out of the traces: indeed, restricting our computation to a spherically symmetric space we again find complex eigenvalues.

Crucially, increasing the dimension of the truncation subspace with respect to the Einstein-Hilbert case adds one UV-attractive and one UV-repulsive eigendirection to the stability matrix, so that the UV critical hypersurface in the truncation subspace is now three-dimensional. We should note that the same number of UV-attractive eigendirections has also been found within the $f(R)$ studies of \cite{Codello:2008vh,Machado:2007ea}, which, however, did not include the power-counting marginal coupling $g_3$.
We then have a three-dimensional subspace of RG trajectories which are attracted to the NGFP in the UV and are therefore ``asymptotically safe''.  Thus, non-perturbative renormalizability persists also in the presence of the 
one-loop perturbative counterterms in the truncation ansatz.

\section{Conclusions}

In this paper we have summarized our recent findings on the non-perturbative RG flow of asymptotically safe quantum gravity in a higher-derivative truncation, including tensorial interactions  in the truncation space. Our main result is the presence of a non-Gaussian fixed point similar to the one observed within previous $f(R)$-type truncations. Despite the four-dimensional truncation space, the number of relevant interactions within the "$R^2 +C^2$" truncation is found to be three. 
Thus, compared to the "$R^2$"-results, extending the truncation subspace has not led to additional relevant directions, providing further evidence for the finite dimensionality of the UV critical surface. Moreover, we found that a NGFP with very similar properties persists when the theory is supplemented with a free scalar field. This result explicitly shows that, contrary to a common worry, the inclusion of perturbatively non-renormalizable counterterms in the truncation subspace of a gravity-matter theory has no qualitative effect on its fixed point structure. In particular, we find no indication that these interactions are fatal to the non-perturbative renormalizability of the theory.


\begin{theacknowledgments}
  D.B.\ thanks the organizers of the XXV Max Born Symposium ``The Planck Scale'', Wroc\l aw, for the invitation. Research at Perimeter Institute is supported in part by the Government of Canada through NSERC and by the Province of Ontario through MRI. P.F.M. is supported by the Netherlands Organization for Scientific Research (NWO) under their VICI program. F.S. is supported by the Deutsche Forschungsgemeinschaft (DFG) within the Emmy-Noether programm (Grant SA/1975 1-1).
\end{theacknowledgments}





\end{document}